\documentclass[conference]{IEEEtran}

\pdfoutput=1

\usepackage{graphicx}
\usepackage{tabularx}
\usepackage{multicol}
\usepackage{color,soul}
\usepackage{capt-of}
\usepackage{mathtools}
\usepackage{url}

\begin{document}

\title{Fluid Passwords - Mitigating the effects of password leaks at the user level}

\author{
\IEEEauthorblockN{Michael Farcasin}
\IEEEauthorblockA{
Oklahoma State University\\
Stillwater, Oklahoma\\
Email: farcasi@cs.okstate.edu
}
\and
\IEEEauthorblockN{Akhileshwar Guli}
\IEEEauthorblockA{
Oklahoma State University\\
Stillwater, Oklahoma\\
Email: akhileg@ostatemail.okstate.edu
}
\and
\IEEEauthorblockN{Eric Chan-Tin}
\IEEEauthorblockA{
Oklahoma State University\\
Stillwater, Oklahoma\\
Email: chantin@cs.okstate.edu}
}

\maketitle

\begin{abstract}
Password leaks have been frequently reported in recent years, with big companies like Sony, Amazon, LinkedIn, and Walmart falling victim to breaches involving the release of customer information. Even though passwords are usually stored in a salted hash, attackers still guess passwords because of insecure password choices and password reuse. However, the adverse effects of a password breach can be mitigated by changing users' passwords. We introduce a simple yet powerful algorithm to reset user account passwords automatically, while still allowing users to authenticate without any additional effort on their part. We implemented our algorithm as a Firefox add-on that automatically resets a user's password when they log in to their account, and stores the new password in the built-in Firefox password manager.
\end{abstract}

\section{Introduction}
Breaches that leak millions of passwords, e.g. LinkedIn~\cite{linkedinwiki, linkedinpr, linkedinconfirm}, Yahoo~\cite{yahooleak1, yahooleak2}, Walmart~\cite{walpwdleak}, and iCloud~\cite{icloudleak} are becoming a daily news event. But while password leaks can lead to both financial loss and loss of privacy~\cite{seciran, appleamazon}, organizations can be slow to detect breaches~\cite{yodaiken2009systems}.

Many methods proposed to protect users from the effects of password leaks, such as OAuth~\cite{yang2013security}, SAuth~\cite{kontaxis2013sauth}, and Two Factor Authentication~\cite{selvarajan2007simple}, require server-side implementation. Despite these solutions, users still have to create strong, unique, memorable passwords and reset their passwords often. But the amount of effort required to create and maintain such passwords leads users to create similar, sometimes the same, weak passwords for multiple accounts, and make only minor changes when required to change their passwords~\cite{farcasin2015we}.

We propose an effort-free, client-side solution to the problem of password theft in the form of a Firefox add-on to automate password resets, which we call \emph{Fluid Passwords}. The idea behind our add-on is: for a given website, after a user logs in, seek out the site's password reset page, generate a secure random password, and update the password for the account. By leveraging Firefox's built-in password manager, we can then store the updated password for use at the next login, all without requiring any interaction from the user. By leveraging Firefox Sync, users can access both their stored passwords and the add-on on any computer they login with, effectively reducing the number of passwords any one person has to remember to the one they use for Firefox Sync. But the Fluid Passwords add-on is not itself a password manager - instead, its primary task is to automatically find a password reset page and reset a password so the user doesn't have to.

We demonstrated the add-on's success on 23 out of 29 websites from Alexa's top 100 websites~\cite{alexa}, and provide suggestions for making it work on the remaining 6 sites in section~\ref{chap:futurework}. Fluid Passwords addresses 3 main problems:

\begin{enumerate}
\item Creating a strong and memorable password is difficult.
\item Creating a unique password for every site is difficult.
\item Changing passwords is difficult, and especially changing passwords often enough to foil attackers who gain access to an old password is difficult. 
\end{enumerate}

Although solutions to the problems exist, client-side solutions to changing passwords are still imperfect and require user interaction~cite{gottLastpass, dashlanePasswordChanger}. In the case of users on a single computer, i.e. users who don't need Firefox Sync, our add-on solves all of these problems. In the case of users on multiple computers, our add-on reduces the problems to the problem of maintaining a single password, i.e. the one to access Firefox Sync. We claim this add-on helps mitigate the effect of password leaks by continuously changing a user's passwords.

Our main contributions are:

\begin{itemize}
\item Creating an algorithm to refresh user passwords that doesn't require user interaction.
\item Creating an algorithm to find a website's password-reset page. 
\end{itemize}

We also wrote scripts to automatically submit password-reset forms and login forms. We implemented these in a Firefox add-on, evaluated its performance, and gave it a comparative rating as an authentication scheme.

The rest of the document is organized as follows. In section~\ref{chap:background}, we review previous literature. In section~\ref{chap:design} we detail our algorithm's design and implementation. In section~\ref{chap:evaluation}, we detail the algorithm's successes and failures. In section~\ref{chap:discussion} we examine the usability and security of the implementation. In section~\ref{chap:futurework} we discuss potential improvements. We conclude in section~\ref{chap:conclusion}.

\section{Background}\label{chap:background}
\subsection{Protection After a Leak}
Past research has focused on helping users create strong passwords, e.g. with PwdHash~\cite{ross2005stronger}, an add-on that turns a simple password into a long and random one, allowing users to create memorable passwords and have good security. In the case of a breach where password hashes are stolen, strong enough passwords resist even attacks by experts~\cite{ur2015measuring}. However, password-strengthening solutions like PwdHash cannot, by design, protect a user once their password is stolen or guessed. In that scenario, a user must reset his or her password to prevent an attacker from using it, and must change it enough that the attacker cannot guess the new one.

Previous work to protect user accounts in the face of a password breach has focused mainly on server-side solutions.

O-Auth~\cite{yang2013security} is a partial solution that provides authenticated parties with only the privileges those users should be allowed. However, there are studies that show that OAuth is intrinsically vulnerable to App Impersonation attack~\cite{hu2014application}.

SAuth~\cite{kontaxis2013sauth} employs authentication synergy among different web services. A user wishing to access service S must also authenticate their account on service V, where S and V are any two web services that are used daily. For example, if a user wants to login into Facebook the protocol redirects user to also authenticate their account on Google. However, users might use the same password for both services S and V, and if not, then it places an additional burden on users as they have to remember two passwords to access a single service, and it does not prevent attackers from accessing all the services available to a legitimate user.

Two-factor authentication (2FA)~\cite{selvarajan2007simple} is an increasingly-popular protocol in which users authenticate with two different forms of authentication. For example, users might have to enter a temporary password/confirmation code provided via SMS or email, or authenticate with a smart card~\cite{yang2008two}. OPass~\cite{sun2012opass} is one such protocol that adds an additional one time key, sent to the user's cell phone at the time of authentication. Another 2FA protocol uses ambient noise to authenticate without user interaction~\cite{karapanos2015sound}. As the authentication requires a separate authentication token in addition to a password, users are safe even in case of a password leak. But because of the inconvenience of waiting for the one-time password, confirmation code, or ambient noise tests, these methods are not commonly used.

\subsection{Protection from Password Leaks}
In addition to securely storing passwords with salted hashes, various algorithms and techniques have been developed to detect password leaks and cracking~\cite{DBLP:journals/corr/ChakrabortyM15, mohammedersatzpasswords}. But these solutions are often complex and costly to deploy and maintain.

Hardware scrambling is another way to avoid password leaks~\cite{scrib}. Trusted hardware such as Scramble S-Crib~\cite{hwscramble} is used to scramble passwords before they are hashed on the server. The server hardware holds the encryption keys that scramble the password, so this hardware is required to do any password guessing. Hence even in case of hash leaks, attackers cannot crack the passwords.

Many websites and organizations follow the practice of password expiration on an interval, e.g. 1 to 3 months, after which the users have to change their existing passwords. However, this approach is unscalable for users~\cite{ur2015added}. As a workaround for this problem, users use similar or the same passwords for multiple accounts and across password changes~\cite{farcasin2015we}.

In contrast to the above solutions, our add-on works on the client-side, so any user can employ it for themselves, it works on multiple sites, and it is effort-free for the user. It is free, and requires no special hardware or reliance on third-party software. The only restriction is that users use the Firefox web browser.

\section{Design \& Implementation}\label{chap:design}
\subsection{General Overview}\label{sec:generaloverview}

\begin{figure}
	\centering
	\includegraphics[scale=0.5]{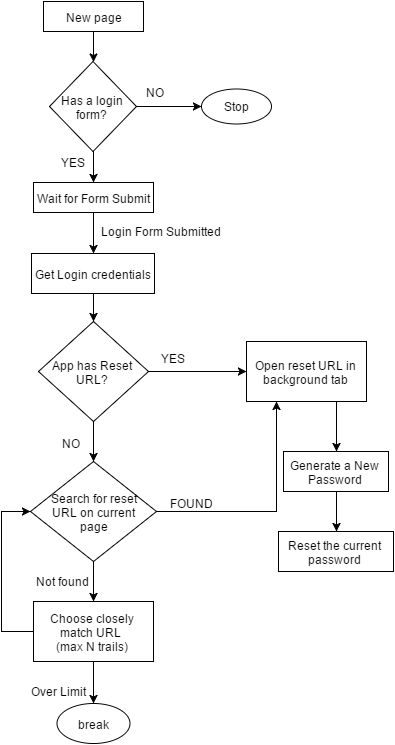}
	\caption{A high-level description of the algorithm.}
	\label{algohighlevel}
\end{figure}

The flow diagram for Fluid Passwords is illustrated in figure~\ref{algohighlevel}. The steps are as follows:
\begin{enumerate}
	\item Identify when the user accesses a login page. When the user submits the login form, go to the next step.
	\item Capture the login credentials. After a successful login, go to the next step.
	\item Search for a link to the site's password reset page.
	\begin{enumerate}
		\item First, check for the site in a file that maps a site's base address to its password-reset page (which we called \textit{purls.txt}). If the site is found, go to the next step.
		\item Otherwise, search the current page for the URL most likely to lead to the password reset page.
	\end{enumerate}
	\item Push the current tab onto a stack and open the selected URL in a new tab.
	\begin{enumerate}
		\item If the new page is a password reset page, generate a new random password and submit the password reset request. Then store the new password in Firefox's password manager. If this site wasn't in purls.txt, store its site/password-reset-page URL pair now.
		\item If the new page is a login page, enter the captured credentials, submit the login form, and return to the previous step.
		\item Otherwise return to the previous step.
	\end{enumerate}
	\item Close any extra tabs opened by the program.
\end{enumerate}

We chose a successful user-login as the start time for our add-on for two reasons. First, it provided a simple way to obtain the user's credentials in case they weren't already stored in Firefox's password manager. Second, it means the add-on runs whenever the user logs in, which we expect to be far more often than the 60-day password expiration guideline published by NIST~\cite{nist:checklist}. 

However, it is important to note that if the add-on cannot find the password reset page, then no password is changed; so the add-on fails safely. In addition, all the websites we observed had only one password reset page, so the add-on will not accidentally change the password saved in Firefox's password manager without actually resetting the password on the site.

\begin{table}
	\small
	\centering
	\begin{tabular}{||c |c||}
		\hline
		Hyper link pattern & Priority level\\
		\hline
		privacy&0\\
		setting&1\\
		profile&2\\
		account&3\\
		security&4\\
		preference&5\\
		my login&6\\
		edit profile&7\\
		password&8\\
		change password&100\\
		\hline
	\end{tabular}
	\caption{Priority and Patterns}
	\label{prioritylevels}
\end{table}

The critical steps are as follows:
\begin{enumerate}
	\item\textbf{Identifying the Login Page} 
	\begin{itemize}
		\item Login pages can almost always be identified by searching for a form with a pattern of text field, password field and a submit button.
	\end{itemize}
	\item\textbf{Capture Login credentials} 
	\begin{itemize}
		\item After identifying the fields above, we simply get the text in those fields.
	\end{itemize}
	\item\textbf{Identifying Successful Login}
	\begin{itemize}
		\item Most of websites redirect to the login page on a failed login attempt; therefore a successful login can be identified by failing to find a login form after entering the credentials.
		\item We can also identify failed attempts by searching for strings that indicate an error, like ``invalid password'', or ``incorrect username''.
	\end{itemize}
	\item\textbf{Selecting a URL} 
	\begin{itemize}
		\item Each URL is assigned a priority value. Priority values are chosen based on the text string that an $<a></a>$ tag encapsulates, with higher priority values given to the strings that are more likely to lead to a password reset page. Then the URL with the highest priority value is chosen. For example, a hyperlink with the text ``password'' has higher priority value than a hyperlink with the text ``account'', so the algorithm would select the URL in the tag with the first string.\footnote{At this time the algorithm has hardcoded English phrases it searches for, though in the future it would be appropriate to include search terms in an easily-updated file, especially one with multiple languages to account for non-English websites.}
		
		The current implementation has 10 priority levels, as shown in table~\ref{prioritylevels}. We were unable to find any literature on the subject, so values were chosen based on our own observations of the websites listed in table~\ref{websiteConsidered}. These were the most common links used to navigate for password reset. For example, on Facebook we navigate from \emph{Settings} $\rightarrow$ \emph{General} $\rightarrow$ \emph{Edit Password}. The further away a link is from the password reset page, the lower priority value it received.
		
		\item The URL with highest priority value on the current page is selected. Then the current tab is pushed onto a stack, and the selected URL is opened in a new tab. We repeat this process until we land on password reset page.
		\item If the new tab does not produce any URL, we move back to the previous tab (popping the current tab from the stack) and select the URL with next highest priority.
	\end{itemize}
	\item\textbf{Opening the URL in a New Tab}
	\begin{itemize}
		\item The add-on opens the new tab in the background. While users can see the tabs that the add-on opens and closes, the tab the user logged in with remains the active tab (assuming the website doesn't change it), so the work in the background tabs does not affect the user's experience.
	\end{itemize}
	\item\textbf{Identifying a Password Reset Page}
	\begin{itemize}
		\item As we were unable to find any literature on identifying password reset pages, we designed our algorithm based on our own observations. Password reset pages have various designs, but most of them have a form with a text field (for usernames), three password fields, then a submit button. We can identify a password reset page by searching for a form with this design.
	\end{itemize}
\end{enumerate}

\subsection{Password Generation}
The contribution of Fluid Passwords is the automated password reset after successful logins and finding the password reset page for each website. Any password generation algorithm can be used to generate a secure
password to be stored in the browser's password manager. We created a simple password generator for the add-on to make it compatible with the websites we tested on, and examine the security of that generator in this section.

The password generator produces pseudorandom 12-character strings from the 95 printable ASCII characters minus the space character, because we found the space character did not work on many websites. In other words, the set of 94 characters [0-9A-Za-z\`\textasciitilde!@\#\$\%\textasciicircum\&$\ast$()\_$+-=$\lbrack\rbrack
$\backslash$
\{\}|;':'',./\textless\textgreater ?].

It uses Stanford JavaScript Crypto Library~\cite{sjcl} as a secure pseudorandom number generator to generate passwords. We chose to create 12-character passwords so the add-on would work with sites that require a maximum password length of 12 characters. Sites with stricter requirements are discussed in section~\ref{chap:futurework}.

Since passwords are chosen probabilistically, we can measure the security of the passwords with Shannon entropy. Low entropy would imply the password distribution is fairly predictable. For example, the entropy of the English alphabet is between 0.6 to 1.3 bits per character. Formula~\ref{eq:6.1} shows that we need a minimum 6.55 bits per character to encode the passwords in binary form. Since we cannot use only part of a bit, we require a minimum 7 bits per character, meaning a 12-character password requires 84 bits to encode. 

\begin{equation}\label{eq:6.1}
\begin{split}
H(X) &= -\sum_{i} \textup{P}(x_i) \textup{log}_b \textup{P}(x_i) \\
	 &= -\sum_{i=1,...,94} \frac{1}{94} \textup{log}_2 \left (  \frac{1}{94}\right ) \\
	 &= 6.5546...
\end{split}
\end{equation}

RFC 4086 recommends 56 bits for higher-than-moderate-security passwords in 2015\footnote{They recommend 49 bits, plus 2/3 of a bit per year due to Moore's law, giving us 56 in 2015.}~\cite{rfc4086}. Our algorithm far exceeds that, but it falls short of the recommended 88 bits for a very-high security password\footnote{75 bits (for a password in 1995) $+ 20 * 2/3$}. We could improve the security by adding non-ASCII characters or lengthening the password. However, using non-ASCII characters would make retyping passwords difficult for those users who wished to do so, and lengthening the passwords would cause some websites to reject them for being too long. Therefore, we chose to use the current algorithm at this time.

\subsection{Implementation}\label{sec:implementation}
Since web browsers are the most common interfaces for accessing online accounts, we chose to implement our algorithm as a Firefox add-on.

For testing, we used a computer with an Intel(R) Core(TM) i7-4510U processor, 8GB RAM. It had a 64-bit Windows 8.1 Operating System, and we developed our add-on with \emph{cfx} for FireFox version 42.0, using Mozilla Firefox Addon-SDK 1.17 and JQuery 1.11.3.

\newpage 
\section{Evaluation}\label{chap:evaluation}
\subsection{Websites}
We tested our add-on with sites chosen from Alexa's top 100 websites in December 2015~\cite{alexa}. A detailed list of all the websites is given in appendix~\ref{app:top100}. Table~\ref{websiteConsidered} lists the websites we chose to test our add-on with, and whether or not the add-on supported these websites. In total, we tested our add-on on 29 of the top 100 sites.

Table~\ref{websiteConsidered} lists the websites we considered. In the case of websites where the add-on didn't work, we give the limitation. Limitations included: being unable to fetch a reset link from a button, and being unable to trigger keyboard event listeners for form inputs when we updated the text in an input box. We address these limitations in section~\ref{chap:futurework}. We note that when the add-on doesn't work, it has no adverse effects on a user's experience, and the websites that the user visits still work. On these sites, the user would have to change their password manually, as if they didn't have the add-on.

We list the 72 websites that we did not choose to test on, along with the reasons, in appendix~\ref{app:websiteNOTconsidered}. Although Google was the most popular site on the list, we chose not to consider it because changing a Google account password causes Android tablet- and smartphone-users to login again with their new password. We ignored websites in a foreign language, since the add-on can only match hyperlink strings against text patterns in English. We ignored websites that ask for security questions as part of a password reset after logging in, as it requires more user-interaction to automate such actions. We ignored torrent and adult websites on ethical grounds, and we ignored websites with no logins, such as news sites, or websites which served as landing pages for other sites, since they do not require users to login.

\begin{table*}
\begin{tabularx}{\linewidth}{||c| X ||}
	\hline
	Limitations (if any) & Websites (count: 29)\\
	\hline
	n/a (add-on works, 23 sites) & Facebook.com(2), Yahoo.com(5), Amazon.com(6), Wikipedia.org(7), Twitter.com(9), Linkedin.com(14), Ebay.com(17), Vk.com(21), Instagram.com(26), Reddit.com(32), Pinterest.com(38), Paypal.com(40), Imgur.com(43), Imbd.com(47), Stackoverflow.com(54), Craigslist.org(56), Ups.com(67), Flickr.com(70), Github.com(72), Indeed.com(75), Fc2.com(82), Nytimes.com(90), Wikia.com(95), Dropbox.com(97) \\
	\hline
	Reset link in a button & Tumblr.com(42), Dailymotion.com(93) \\
	\hline
	Unable to trigger event listeners & Wordpress.com(39), Netflix.com(52) \\
	\hline
\end{tabularx}
\caption{A list of websites considered, together with their rank}
\label{websiteConsidered}
\end{table*}

Table~\ref{cpumemoryusage} lists the percent CPU usage and memory usage (in MB) by Firefox web browser during the process of password reset, with and without the add-on enabled. The first two columns are when the add-on was enabled, and the third and fourth columns are when the add-on was not enabled. The last two columns show the difference in overhead caused by the add-on. Overall, the difference is negligible, with the add-on adding only 1.48\% more CPU usage and 16.54 MB more memory usage on average. And it is worth noting that these times were tested the first time we accessed a site. Because the add-on stores the password reset URL and in future logins can directly open the password reset page instead of searching again, we expect it to use even less overhead during future logins.

\begin{table*}
\centering
	\begin{tabular}{||c| c| c| c |c| c |c||}
		\hline
		&\multicolumn{2}{c|}{With Add-on} & \multicolumn{2}{c|}{Without Add-on}	& \multicolumn{2}{c||}{Difference}
		\\
		\hline
		Websites &  CPU\% & Memory(MB) &  CPU\% & Memory(MB) & CPU\%	& Memory(MB)
		\\ [0.5ex]
		\hline
		Facebook & 11.11 & 117.95 & 10.37 & 108.56&0.74&9.9\\
		Yahoo & 10.72 & 156.3& 10.19 & 140.32&0.53&15.98\\
		Amazon & 12.56 & 197.76& 11.74 & 160.52&0.82&37.11\\
		Wikipedia & 9.51 & 130.41& 9.20 & 125.65&0.31&4.76\\
		Twitter & 10.92 & 162.87& 10.81 & 143.51&0.11&19.36\\
		LinkedIn & 8.61 & 125.62& 8.23 & 122.18&0.38&3.44\\
		eBay & 17.80 & 219.86& 13.76 & 201.65&4.04&18.21\\
		Vk.com & 11.58 & 180.18& 11.42 & 160.35&0.16&19.83\\
		Instagram & 15.64 & 156.3& 13.19 & 140.32&2.45&15.98\\
		Reddit & 13.52 & 150.3& 10.72 & 133.32&2.8&16.98\\
		Pinterest & 15.21 & 114.3& 9.46 & 111.20&5.75&3.1\\
		Paypal & 11.43 & 235.27& 10.54 & 200.32&0.89&34.95\\
		Tumblr & 12.61 & 152.40& 12.03 & 141.03&0.58&11.37\\
		Imgur& 13.71 & 112.53& 10.90 & 86.52&2.81&26.01\\
		Fc2 & 10.63 & 113.51& 10.09 &110.3 &0.54&3.21\\
		Stackoverflow & 13.92 & 125.52& 11.57 & 98.51&2.35&27.01\\
		Craigslist & 9.34 & 120.47& 8.68 & 115.8&0.66&4.67\\
		Ups & 14.61 & 130.12& 13.18 & 119.37&1.43&10.75\\
		Flickr & 10.51 & 162.3& 9.43 & 153.4&1.08&8.9\\
		Github & 12.86 & 184.81& 10.96 & 176.31&1.9&8.5\\
		Indeed & 11.89 & 96.53& 10.34 & 92.04&1.55&4.49\\
		Amazon.in & 13.5 & 214.50& 11.57 & 140.32&1.93&74.18\\
		Nytimes & 16.55 & 156.61& 15.81 & 143.12&0.74&13.49\\		
		Wikia & 12.75 & 96.3& 11.63 & 91.32&1.12&4.98\\
		\hline
	\end{tabular}
	\caption{CPU \& Memory Usage}
	\label{cpumemoryusage}
\end{table*}

The add-on's listeners take a few seconds to attach to the web page elements, and they must wait until the web page loads all its content. When the add-on's listeners fail to attach, or the user submits the form before the listeners are attached\footnote{In practice, this happened when a user's password manager auto-filled their credentials and they clicked ``login'' as soon as it appeared}, the add-on does not continue executing: it neither opens any tabs in the background nor resets the user's password. This does not cause the browser to crash, but merely to behave as if the add-on was not installed during this login.

\begin{table}
	\small
	\centering
	\begin{tabular}{|| c |c |c||}
		\hline
		Website&Without Add-on&With Add-on\\
		\hline
		eBay.com&6 sec (avg)&15 sec (avg)\\
		\hline
	\end{tabular}
	\caption{Average time to sign-in to a website.}
	\label{signupdelay}
\end{table}

\subsection{Improvements \& Accuracy}\label{sec:improvementsandaccuracy}
In this section, we discuss the various improvements we made as we went through different versions, making the add-on support websites with different designs. Depending upon the basic flow from a login page to a password reset page, websites can be broadly classified into three categories.

\subsubsection{Simple Websites}\label{simplewebsites}
\begin{figure}
	\centering
	\includegraphics[width=\columnwidth]{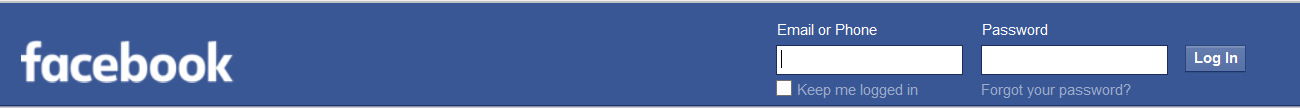}
	\caption{Login on landing page}
	\label{fblogin}
\end{figure}

\begin{figure}
	\centering
	\includegraphics[scale=0.5]{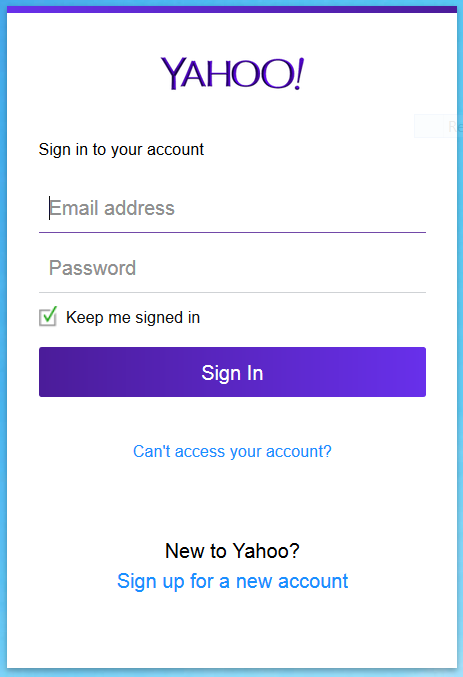}
	\caption{Simple Login page}
	\label{yahoologin}
\end{figure}

\begin{figure}
	\centering
	\includegraphics[width=\columnwidth]{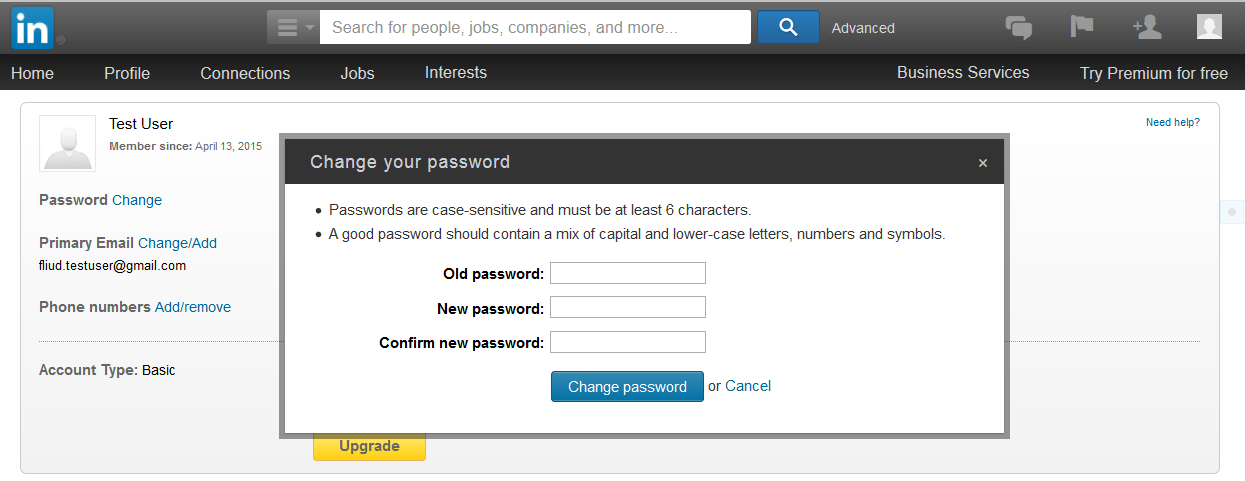}
	\caption{Simple Password reset page}
	\label{linkedinpwdreset}
\end{figure}

These websites have a simple design where the landing page/home page has a basic login form with a user name, a password and a submit button. The design of the password reset page is also simple with an optional username, three password fields, and a submit button. Figures~\ref{fblogin} and~\ref{yahoologin} show the design for simple login pages. Facebook has a login page on its landing page, whereas Yahoo redirects to a new page with simple login design. Figure~\ref{linkedinpwdreset} shows a basic password reset page from Linkedin. The control flow for this website, identifying the login page to submitting the password reset form, is straightforward. 30.7\% of the websites in table~\ref{websiteConsidered} fall under this category. Our initial implementation worked only with these types of websites.


\subsubsection{Websites that require re-login}\label{reloginwebsites}
These websites ask a user to login again before they can change their password. 19.3\% of the websites in table~\ref{websiteConsidered} fall under this category, including Amazon and eBay. To support these websites, we added an additional check for a new login page and added code to wait on the redirect to password reset page.

\begin{figure}
	\centering
	\includegraphics[scale=0.5]{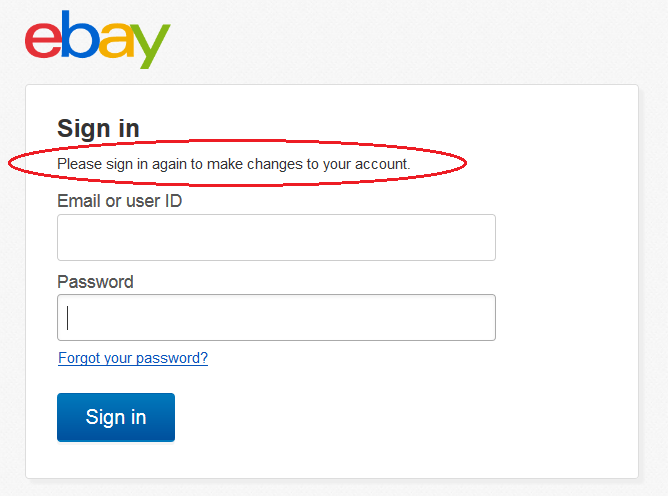}
	\caption{eBay Re-Login page}
	\label{reloginpage}
\end{figure}

\subsubsection{Complex Websites}\label{complexwebsites}
\begin{figure*}
	\centering	\includegraphics[scale=0.5]{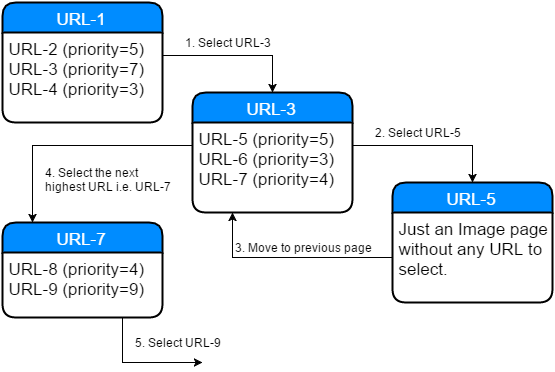}
	\caption{Depth-First Search Design}
	\label{depthfirstdesign}
\begin{enumerate}
\item Select URL-3 (the highest priority URL) and open it in a new tab.
\item The new tab with URL-3 has 3 URLs. URL-5 has the highest priority level, so open URL-5 in a new tab.
\item The new tab has no new URLs, so we go back to the previous page.
\item URL-7 has the next highest priority, so we open it in new tab.
\item URL-9 has the highest priority in this tab, and the process continues.
\end{enumerate}
\end{figure*}

We chose priority values for URLs based on our observation that the URL would lead to password reset page. Therefore, it's not always certain that we will find the password reset page. The initial algorithm design would stop when both the current page was not a password reset page, and the search process found no new links. But e-commerce sites such as Amazon and eBay have a large number of hyperlinks, and in fact 50\% of the websites listed in table~\ref{websiteConsidered} fall under this category. To handle these scenarios, we implemented a depth-first search, in which failure causes the algorithm to go back to the previous tab and select the URL with the next-highest priority. In theory, this design would exhaustively search a site's URLs and eventually find the password reset page. However, we set a threshold of 20 opened pages, after which the algorithm terminates. Figure~\ref{depthfirstdesign} shows how this might work in practice. 

\subsection{Limitations}\label{sec:limitations}
The add-on does not support websites that send a confirmation code or reset link to user's email/phone, or that ask for security questions in order to reset a password. And many banking websites require users to enter a CAPTCHA to login, so our add-on does not support those websites when a re-login is required.

In addition, the current implementation does not support any Google-based website, because Google's password reset page leverages a div tag for form submission.

On the other hand, Wordpress and Netflix both listen for actual keyboard events before allowing form submission, so simply assigning a value to a text box with textBox.value does not work in this case.

We provide suggestions for addressing these limitations in section~\ref{chap:futurework}.

\section{Discussion}\label{chap:discussion}
\subsection{Comparative Rating}\label{sec:rating}
We rate our add-on using Bonneau's framework for comparative evaluation of authentication schemes~\cite{bonneau2012quest}. Naturally, our scheme inherits many of its qualities from passwords and Firefox's password manager, though in some cases it exceeds both. 

\textbf{Usability}
\begin{enumerate}
\item It is \emph{memorywise-effortless} for users at a single computer, and quasi-memorywise-effortless for users of Firefox Sync, as the sync password is the only password users need to remember.
\item It is \emph{scalable-for-users}, as the add-on and password manager handle all the passwords.
\item It is \emph{nothing-to-carry}.
\item It is \emph{quasi-physically-effortless}, for the same reasons Firefox's password manager is.
\item It is \emph{easy-to-learn}, because passwords are.
\item It is \emph{efficient-to-use}, more so than passwords because this does not require typing.
\item The current version does not have \emph{infrequent-errors}. The ideal version (that works on every website) would have \emph{infrequent-errors}, as users need only click the login button.
\item It does not have \emph{easy-recovery-from-loss}, for the same reasons Firefox's password manager doesn't.
\end{enumerate}

\textbf{Deployability}
\begin{enumerate}
\item It is \emph{accessible}, for the same reasons Firefox's password manager is.
\item It has \emph{negligible-cost-per-user}.
\item The current version is \emph{quasi-server-compatible}, as it works on some websites but not others. The ideal version would be \emph{server-compatible}.
\item It is not \emph{browser-compatible}, though in future work we would extend the add-on to other browsers.
\item It is not \emph{mature}, as it is newly-created.
\item It is \emph{non-proprietary}.
\end{enumerate}

\textbf{Security}
\begin{enumerate}
\item It is \emph{quasi-resilient-to-physical-observation}, like Firefox's password manager.
\item Likewise it is \emph{quasi-resilient-to-targeted-impersonation}.
\item It is \emph{resilient-to-throttled-guessing}, as it generates secure, random passwords.
\item It is \emph{resilient-to-unthrottled-guessing}, as there are $94^{12} \approx 2^{78}$ possible passwords.
\item It is not \emph{resilient-to-internal-observation}.
\item We claim it is \emph{quasi-resilient-to-leaks-from-other-verifiers}, both because it generates unique passwords for each site and because it resets passwords relatively quickly.
\item It is \emph{resilient-to-phishing}.
\item It is \emph{resilient-to-theft} under the same assumptions that Firefox's password manager is.
\item It does not offer \emph{no-trusted-third-party}.
\item It \emph{requires-explicit-consent}.
\item It is \emph{unlinkable}, because it generates random passwords.
\end{enumerate}

\subsection{Firefox Sync}

Firefox Sync is a service offered by Firefox wherein users can create an account on Firefox that allows them to share information across devices, especially passwords and add-ons~\cite{firefoxSync}. Users login to their account with a username and password, so Firefox Sync acts as an online password manager with a master password.

When users have Fluid Passwords without using Firefox Sync, they are limited to storing their passwords on one device, so that Fluid Passwords will not function on other devices (because it won't be able to access the user's login credentials). However, when users employ both Fluid Passwords and Firefox Sync, and share passwords and add-ons across devices, Fluid Passwords will function any time the user is using Firefox, and their updated passwords will be stored and accessible on any of their devices running Firefox.

\subsection{Other considerations}\label{sec:otherconsid}
The add-on is very simple to use, as it only requires the user to run Firefox and install the add-on. However, the current version of the add-on only works on websites for which the user has stored his or her credentials in Firefox's password manager. For example, if a user stores their Facebook credentials in the password manager, but not their Linkedin credentials, then the add-on would only work with their Facebook account, not Linkedin. Similarly, the add-on doesn't work for multiple users. For example, if two users, A and B, both use the add-on, but A stores their Facebook password in the password manager and B doesn't, then the add-on would only change A's password. And Fluid Passwords may cause problems for users who use the add-on on one device, e.g. their computer, but also need to login to a site with another device that doesn't have access to the add-on, e.g. an iPhone. With the current version of the add-on, the only way for such a user to login with their second device would be to look up their password in Firefox's password manager, or on Firefox Sync if they use it. However, because Fluid Passwords relies on Firefox's password manager, the cost of losing a password is as low as using the password manager alone. The user would need to reset their password and store the new password in their password manager.

Once the add-on has successfully reset a password on a site, it stores the URL of the password reset page in a separate file named \textit{purls.txt}. After that, whenever it encounters the same website again, it uses the link in \textit{purls.txt} to go directly to password reset page. This has 3 main implications:
\begin{enumerate}
\item Users can share \textit{purls.txt} files to increase our pool of url mappings. In particular, users could submit their \textit{purls.txt} files to a central server, which could then update a master list and distribute it to users, effectively crowdsourcing password-reset-URL searching.
\item Sites that have personalized URLs for their password-reset pages cannot be usefully stored in purls.txt, and the add-on will be forced to search for the URL each time the user logs in.
\item It allows the add-on to work on non-English sites. Any site in the \textit{purls.txt} file ignores the (currently) English-only url-searching algorithm and jumps directly to the password reset page, where it looks only at form element types, which are all written in HTML language.
\end{enumerate}

We must also discuss the insecurity of an add-on that automatically fills in user credentials when it finds a login form. Once the user triggers the add-on's automatic password reset functions by logging in, it may navigate to sites with malicious scripts that request login credentials to steal them. We don't have a good solution for this problem, except to rely on the sites themselves to keep the pages a user sees after logging in free of such scripts.

\subsection{Future Work}\label{chap:futurework}
Capturing login credentials is difficult for websites where the web page has dynamic forms, forms nested in iframes, and other complexities\footnote{e.g. zillow.com, etsy.com, buzzfeed.com}. Identifying a successful login becomes difficult when we have hidden or dummy login-forms\footnote{e.g. godaddy.com has a dummy login form on the user's profile page}, or when it is difficult to identify a login page\footnote{e.g. aliexpress.com, alibaba.com, flipkart.com}.

Traversing URLs is the next crucial task to reach password reset page. Our current version looks for text hyperlinks on the page to assign priority, but this method fails when the URLs are embedded as hyperlinks in images or buttons. In addition, identifying a password reset page is difficult when forms on the page do not follow the normal pattern\footnote{Optional username, three password fields, then a submit button. E.g. wordpress.com, themeforest.net}. Finding a single algorithm for all sites on the web would be an important step forward.

Two sites stand out in particular: Google and Dropbox. Google's password reset page has a div tag with a click-listener for its form submission. Assuming the tag has the same id for every user, we might attempt to find the tag and issue it a JavaScript click. Dropbox listens for keyboard events to fill in its password reset form, denying our script the ability to do so, and the tags in its forms have enough unique information to prevent our simply creating our own input tags with the required information. We might try simulating keyboard events with JavaScript to fill in such forms.

While our primary concern is getting our add-on to work on every site on the web, we would ideally also extend our add-on to Google Chrome, Safari, and other browsers, as well as mobile phones. The add-on currently does not mesh well with mobile phone usage, since mobile phones typically do not have the same add-ons installed as desktop computers. However, Firefox and Chrome at least allow mobile phone users to sync their passwords across devices, so would have up-to-date passwords on their mobile phones.

We would also like to add the ability for users to blacklist sites that they do not want the add-on to function on. This would allow users to handle passwords for those sites on their own, while still gaining the benefit of the add-on on other sites. And such a blacklist could be seeded with sites that require passwords shorter than 12 characters, so that the add-on doesn't waste time attempting to create a new password for that site every time a user logs in. We note, however, that the add-on does not break on any site - it simply exits without changing a user's password and closes any open tabs, keeping the process invisible to the user.

The code for generating passwords currently only produces random 12-character passwords, but some sites have more exact requirements, e.g. at least 1 uppercase, 1 lowercase, 1 number, and 1 symbol. Future work for the add-on would include detecting a site's password requirements and modifying the password generation algorithm to create passwords that fit them.

Last, since the goal of Fluid Passwords is to provide better usability and security, a user study would confirm our tool is user-friendly and non-intrusive, but we expect that to be the case due to the design of the algorithm. We stress that this add-on works in the background, so it does not interfere with a user's browsing experience.

\section{Conclusion}\label{chap:conclusion}
We developed an algorithm to automatically reset user passwords and implemented it in \emph{Fluid Passwords}, a Firefox add-on that leverages Firefox's password manager to improve the security and usability of passwords online. Fluid Passwords solves the problem of password reuse, and reduces the problems low password strength, and the difficulty in making frequent password changes to a single password, allowing users to put more focus on maintaining a secure password while expending less effort overall. In its current version, it works on 23 out of 29 sites from Alexa's top 100 sites, and has inherent functionality to work on many more, and adds negligible cost to performance with only a small cost to login-time.

While there are many more improvements that can be made, this is currently the only client-side solution we know of to keep accounts secure after a password leak without user interaction. We plan to release Fluid Passwords as an official Firefox add-on.

\section{Acknowledgments}
Thanks to the contributors of Mozilla's Addon-SDK tutorials and API, and thanks to Dr.-Ing. Roland Bless for creating a bibtex citation of RFC 4086.





\newpage
\appendix
\section{Top 100 Websites}\label{app:top100}
\begin{multicols}{2}
\begin{enumerate}
\item Google.com
\item Instagram.com
\item Google.es
\item Indeeed
\item Facebook.com
\item Msn.com
\item Googleadservices.com
\item Cnn.com
\item Youtube.com
\item Microsoft.com
\item Netflix.com
\item Amazon.in
\item Baidu.com
\item Aliexpress.com
\item Amazon.de
\item Go.com
\item Yahoo.com
\item Amazon.co.jp
\item Stackoverflow.com
\item Google.co.id
\item Amazon.com
\item Google.co.uk
\item 360.cn
\item Xinhuanet.com
\item Wikipedia.org
\item Reddit.com
\item Craigslist.org
\item Blogger.com
\item Qq.com
\item Ask.com
\item Tianya.cn
\item Google.com.au
\item Twitter.com
\item Google.fr
\item Diply.com
\item nytimes
\item Google.co.in
\item Google.com.br
\item Ok.ru
\item Bbc.co.uk
\item Taobao.com
\item Tmall.com
\item Google.ca
\item People.com.cn
\item Live.com
\item Onclickads.net
\item Alibaba.com
\item Cntv.cn
\item Sina.com.cn
\item Pinterest.com
\item Google.com.mx
\item Pixnet.net
\item Linkedin.com
\item Wordpress.com
\item Pornhub.com
\item Gmw.cn
\item Yahoo.co.jp
\item Paypal.com
\item Google.com.hk
\item Ebay.de
\item Weibo.com
\item Mail.ru
\item Naver.com
\item Google.pl
\item Ebay.com
\item Tumblr.com
\item Amazon.co.uk
\item Googleusercontent.com
\item Google.co.jp
\item Imgur.com
\item Ups
\item Dailymotion.com
\item Yandex.ru
\item Sohu.com
\item Xhamster.com
\item Google.co.kr
\item Blogspot.com
\item Xvideos.com
\item Rakuten.co.jp
\item Wikia.com
\item Vk.com
\item Google.ru
\item flickr.com
\item Chinadaily.com.cn
\item Hao123.com
\item Imdb.com
\item Kat.cr
\item Dropbox.com
\item T.co
\item Apple.com
\item github.com
\item Livedoor.jp
\item Bing.com
\item Google.it
\item Soso.com
\item Ebay.co.uk
\item Google.de
\item Fc2.com
\item Nicovideo.jp
\item Dailymail.co.uk
\end{enumerate}
\end{multicols}

\newpage
\section*{Websites Not Considered for Testing}\label{app:websiteNOTconsidered}
\noindent\begin{minipage}{\textwidth}
\begin{tabularx}{\linewidth}{|| c | X ||}
	\hline
	Reason	&	Websites (count: 71)\\
	\hline
	App breaks the live smartphone connection & Google.com(1)\\
	\hline
	No Login(2)	&	Ask.com(33), Cnn.com(76)\\
	\hline
	Torrent(1)	&	Kat.cr(71)\\
	\hline
	Adult(3) 	&	Xvideos.com(45), Pornhub.com(63), Xhamster.com(68)\\
	\hline
	Landing page for other sites(2)	&	Go.com(78), Onclickads.net(37)\\
	\hline
	Security Question(7)	&	Live.com(12), Bing.com(24), Msn.com(27), Microsoft.com(28), Aliexpress.com(29), Apple.com(48), Alibaba.com(61)\\
	\hline
	Duplicate(12) & Youtube.com(3), Google.co.in(10), Blogspot.com(20), Google.co.uk(31), Googleadservices.com(51), Google.ca(60), Blogger.com(81), Google.com.au(83), Googleusercontent.com(92), Amazon.co.uk(66), Amazon.in(77), Ebay.co.uk(99)\\
	\hline
	Foreign website(44)	&	Baidu.com(4), Qq.com(8), Taobao.com(11), Sina.com.cn(13), Yahoo.com.jp(15), Weibo.com(16), Google.co.jp(18), Yandex.ru(19), Hao123.com(22), T.co(23), Google.de(25), Amazon.co.jp(30), Google.fr(34), Google.com.br(35), Tmall.com(36), Mail.ru(41), Sohu.com(44), Google.ru(46), Google.it(49), Google.es(50), Amazon.de(53), 360.cn(55), Tianya.cn(57), Diply.com(58), Ok.ru(59), Google.com.mx(62), Google.com.hk(64), Naver.com(65), Rakuten.co.jp(69), Soso.com(73), Nicovideo.jp(74), Google.co.id(79), Xinhuanet.com(80), Bbc.co.uk(84), Pople.com.cn(85), Cntv.cn(86), Pixnet.net(87), Gmw.cn(88), Ebay.de(89), Google.pl(91), Google.co.kr(94), Chinadaily.com.cn(96), Livdoor.jp(98), Dailymail.co.uk(100)\\
	\hline
\end{tabularx}
\captionof{table}{A list of websites not considered, together with their rank}
\label{websiteNOTconsidered}
\end{minipage}
\end{document}